\documentstyle[12pt,aaspp4,flushrt,epsfig,natbib]{article}	% nice
\bibpunct{(}{)}{;}{a}{}{;}	% natbib

\newcommand{\etal}{\textit{et~al.}}
\newcommand{\hmpc}{\mbox{$\:h^{-1}\:\mbox{Mpc}$}}
\newcommand{\twod}{two-dimensional}
\newcommand{\tred}{three-dimensional}
\newcommand{\flux}{\mbox{${\cal F}$}}  % energy flux
\newcommand{\lum} {\mbox{${\cal L}$}}  % luminosity
\newcommand{\centercol}[1]{\multicolumn{1}{c}{#1}}
\newcommand{\sect}[1]{\mbox{\S \ref{#1}}}
\newcommand{\Fig}[1] {Fig.~\ref{#1}}
\newcommand{\eq}[1]  {eq.~[\ref{#1}]}
\newcommand{\equ}[1] {equation~(\ref{#1})}
\newcommand{\beeq}{\begin{equation}}
\newcommand{\eneq}{\end{equation}}
\newcommand{\astroref}[1]{\texttt{(astro-ph/#1)}}
\newcommand{\definition}[1]{{\sffamily \slshape #1}}

\newcommand{\hour}  {\mbox{$^{\rm h}$}}            % hours   (^h)
\newcommand{\Min}   {\mbox{$^{\rm m}$}}            % minutes (^m)
\newcommand{\phmh}  {\phantom{\hour}}       % hour
\newcommand{\phmm}  {\phantom{\Min}}        % minute
\newcommand{\phmdg} {\phantom{\arcdeg}}     % arcdeg
\newcommand{\phmmin}{\phantom{\arcmin}}     % arcmin
\newcommand{\phmgt} {\phantom{\mbox{$>$}}}    % greater than

\newcommand{\iras}{\textit{IRAS}}
\newcommand{\bootes}{Bo\"otes}
\newcommand{\potent}{\textsc{potent}}
\newcommand{\wb}{\textsc{wall builder}}
\newcommand{\vf}{\textsc{void finder}}
\newcommand{\rmax}{\mbox{$r_{\rm max}$}}
\newcommand{\dmax}{\mbox{$d_{\rm max}$}}
\newcommand{\rvol}{\mbox{$r_o$}}
\newcommand{\di}{\mbox{$d_{\tt i}$}}
\newcommand{\dstop}{\mbox{$d_{\rm stop}$}}
\newcommand{\npoisson}{\mbox{$N_{\rm Poisson}(d)$}}
\newcommand{\nsurvey}{\mbox{$N_{\rm Survey}(d)$}}

\lefthead{El-Ad \& Piran}
\righthead{Voids in the Large-Scale Structure}

\slugcomment{Submitted to ApJ}

\begin{document}

\title{Voids in the Large-Scale Structure}

\author{Hagai El-Ad\authoremail{eladh@shemesh.fiz.huji.ac.il} and
        Tsvi Piran\authoremail{tsvi@shemesh.fiz.huji.ac.il}}
\affil{Racah Institute for Physics, The Hebrew University, Jerusalem, 91904 Israel}

\begin{abstract}
Voids are the most prominent feature of the large-scale structure of the 
universe. Still, they have been generally ignored in quantitative analysis of 
it, essentially due to the lack of an objective tool to identify the voids 
and to quantify them. To overcome this, we present here the \vf\ algorithm, a 
novel tool for objectively quantifying voids in the galaxy distribution. The 
algorithm first classifies galaxies as either wall galaxies or field 
galaxies. Then, it identifies voids in the wall galaxy distribution. Voids 
are defined as continuous volumes that do not contain any wall galaxies. The 
voids must be thicker than an adjustable limit, which is refined in 
successive iterations. In this way we identify the same regions that would be 
recognized as voids by the eye. Small breaches in the walls are ignored, 
avoiding artificial connections between neighboring voids.

We test the algorithm using Voronoi tessellations. By appropriate scaling of 
the parameters with the selection-function we apply it to two redshift 
surveys, the dense SSRS2 and the full-sky \iras\ 1.2~Jy. Both surveys show 
similar properties: $\sim 50\%$ of the volume is filled by the voids. The 
voids have a scale of at least 40\hmpc, and an average under-density of 
$-0.9$. Faint galaxies do not fill the voids, but they do populate them more 
than bright ones. These results suggest that both optically and \iras\ 
selected galaxies delineate the same large-scale structure. Comparison with 
the recovered mass distribution further suggests that the observed voids in 
the galaxy distribution correspond well to under-dense regions in the mass 
distribution. This confirms the gravitational origin of the voids.
\end{abstract}


%  ****************************************
%  *        Section 1: Introduction       *
%  ****************************************

\section{Introduction}

Perhaps one of the most intriguing findings of dense and complete nearby 
redshift surveys has been the discovery of large voids on scales of 
$ \sim 50 \hmpc $, and that such large voids appear to be a common feature of 
the galaxy distribution. Early redshift surveys like the Coma/A1367 survey 
\citep{gt78} and the Hercules/A2199 survey \citep{crt81} gave the first 
indications for the existence of voids, each revealing a void with a diameter 
of $ \sim 20 \hmpc $. Surprising as these findings may have been, it was not 
before the discovery of the \bootes\ void \citep{kir81} that the voids caught 
the attention of the astrophysical community (for a review, see 
\citealt{rood88}).

The unexpectedly large void found in the \bootes\ constellation, confirmed to 
have a diameter of $ \sim 60 \hmpc $ \citep{kir87}, brought up the question 
whether the empty regions we observe are a regular feature of the 
distribution of galaxies, or rather rare exceptions. Wide-angle yet dense 
surveys, initially \twod\ and more recently \tred, probing relatively large 
volumes of the nearby universe, established that the voids are indeed a 
common feature of the large-scale structure (LSS) of the universe. The 
publication of the first slice from the CfA redshift survey \citep{lap86} 
introduced the picture of a universe where the galaxies are located on the 
surfaces of bubble-like structures, with diameters in the range 25--50\hmpc. 
The extensions of the CfA survey \citep{gh89}, complemented in the south 
hemisphere by the SSRS and its extension, the SSRS2 \citep{dc88,dc94} have 
shown that not only large voids exist, but more importantly---that they occur 
frequently (at least judging by eye), suggesting a compact network of voids 
filling the entire volume.

It has been recognized early on that inhomogeneities on such scales could 
impose strong constraints on theoretical models for the formation of LSS\@. 
However, the voids have been largely ignored and their incorporation into our 
theories of LSS has been relatively recent \citep{blu92,du93,pi93}\@.
The major obstacle confronting the incorporation of voids into models
of LSS has been the difficulty of developing proper tools to identify 
and to quantify them in an objective manner.
As such, the description of a void-filled universe with a characteristic scale
of 25--50\hmpc\ relied solely on
the visual impression of redshift maps. In order to make a more
quantitative analysis we have developed an algorithm (\vf) for the automatic
detection of voids in \tred\ surveys. 
Unlike other statistical measures (e.g., the VPF), our target is to identify
the \emph{individual voids}, in as much the same way as voids are identified 
by eye. The main features of the algorithm are: 
\begin{enumerate}
\item
It is based on the point-distribution of galaxies, without introducing 
any smoothing scale which destroys the sharpness of the observed features.
\item
It allows for the existence of some galaxies within the voids, recognizing 
that voids need not be completely empty.
\item
It attempts to avoid the artificial connection between neighboring voids
through small breaches in the walls, realizing that walls in the galaxy
distribution need not be homogeneous as small-scale clustering will
always be present.
\end{enumerate}

After a review of earlier works (\sect{early}), we present a detailed 
description of our algorithm (\sect{algo}). We use Voronoi tessellations as a 
test bed for the algorithm in \sect{voro}. We then apply the algorithm to the 
SSRS2 redshift survey (\sect{ssrs2}). Finally (\sect{discuss}), we compare the 
voids found in SSRS2 and in \iras\ 1.2~Jy, and discuss the results.

%  ****************************************
%  *       Section 2: Earlier Works       *
%  ****************************************

\section{Earlier Works}
\label{early}

The various methods for describing the void content of the LSS of the 
universe can be divided into two categories: statistical measures, and
algorithms for identifying individual voids within a sample.

The major statistical tool used for describing the voids is the 
\emph{Void Probability Function} (VPF). It measures the probability $P_0(V)$
that a randomly positioned sphere of volume $V$ contains no galaxies
\citep{wt79}.
For a completely uncorrelated (Poissonian) distribution, it is:
\beeq P_0(V) = \exp(-nV) \eneq
where $n$ is the number-density of galaxies, so
that any departure from this quantity represents the signature for
the presence of clustering.
The major drawback of the VPF is that it is very sensitive to the
details of the galaxy distribution. For instance, adding a few galaxies in
the under-dense regions may greatly modify the VPF. A less sensitive
variant of the VPF is the \emph{Under-dense Probability Function} (UPF),
defined as the probability $ P_{\delta \rho / \rho}(V) $ that a randomly 
positioned sphere of volume $V$ has a $ \delta \rho / \rho $ under-density 
\citep{lw94}.

The first zero-crossing of the two-point correlation function $\xi(r)$ was 
used by \cite{gol95} for determining the maximum diameters of voids in case
of spherical voids. The two-point correlation function is defined as the
probability in excess of Poisson distribution of finding a galaxy in a 
volume $\delta V$ at a distance $r$ away from a randomly chosen galaxy:
\beeq  \delta P = n \, \delta V \, [ 1 + \xi(r) ]  \eneq 
where $n$ is the mean galaxy number-density. For a cellular like distribution 
the first-zero crossing is a direct measure of the characteristic size of the 
cells. Using Voronoi tessellations, they show that despite the large 
uncertainty in the determination of $\xi(r)$ on large scales ($ > 20 \hmpc $), 
the zero-crossing statistic may be a useful tool in determining the scale of 
typical voids---if the galaxy distribution is void-filled, and if there is a 
characteristic scale for the void distribution. Both these questions are 
addressed here. Examining the SSRS2 sample, it was found that 
$ R_{\rm zero} \approx 38 \hmpc $.

Previous works have used various definitions for voids, and applied 
different algorithms to identify them.
Perhaps the first work identifying voids in a quantitative manner is
that of \cite{pel89}, who examined ensembles of contiguous cells 
with densities below a given threshold.
They use a cubic lattice, and define a local density
for each cell in the lattice. This local density is based on the analysis of
the smoothed density field, constructed from the original discrete
galaxy distribution. Groups of cells with densities below a specific
limit constitute the voids. The algorithm considers
two cells as contiguous if they are in contact either through their
faces, edges or vertices. This technique was applied to the original
SSRS, identifying 4 to 8 voids, depending on the density threshold. 
Unfortunately, there is only partial overlap between the SSRS and the SSRS2.
Thus, an overall comparison between the results is impossible (where possible, 
corresponding voids are indicated in Table~\ref{ssrs2-table}). 
The major shortcomings of this algorithm are its use of a smoothed density 
field, and the lack of sense of the shape of the void it recognizes,
allowing for practically any void shape.

\citeauthor{kau91} (\citeyear{kau91}, hereafter KF91) designed a more 
elaborate algorithm. They too used (empty) cubes, to which adjacent faces are 
attached. However, in order to avoid long finger-like extensions leading from 
one void into other voids, they impose a constraint on the adjacent faces, 
that each face must have an area of no less than two-thirds that of the 
surface on to which it was to be added. This scheme is restrictive, as it is 
tailored for finding only ellipsoidal-shaped voids. The algorithm was applied 
to the \emph{Southern Redshifts Catalog} (SRC) and to an all-sky catalog, 
finding a peak in the spectrum of void diameters between 8 and 11\hmpc.

Only $\sim 10$ (out of $> 100$) of the voids found in KF91 are located within 
the boundaries of the SSRS2. About half of the KF91 voids have an equivalent 
diameter $ d < 20 \hmpc $, which we believe to be statistically insignificant 
(see \sect{stat}). Other KF91 voids are much smaller than their counterparts 
identified by the \vf: the Sculptor void (void~5 here) is 40\% larger in our 
analysis, and other voids we find are up to three times larger than the 
corresponding KF91 voids. The limited algorithm and the different statistical 
analysis have resulted in a small void diameter, not describing the true 
nature of the void distribution.

%  # KF91       r   r.a.  dec  vol(K)   VF   r   r.a.  dec  vol(K) RELATION
%  1 Sculptor  50.6 23.6 -33.6  16       5  53.0 23.8 -24.7  22.6    *1.4
%  2        2   7.5  2.2 -15.5   0.06   --  ---- ---- -----  ----    TINY
%  3        5  19.1 21.0 -35.3   0.46   --  ---- ---- -----  ----    TINY
%  4       13  11.5 23.0  -5.0   0.36   --  ---- ---- -----  ----    TINY
%  5       22  26.3  4.1 -17.7   0.90   --  ---- ---- -----  ----    TINY
%  6       32  33.8  4.5 -30.2   1.3    17  32.1  3.7 -21.4   3.7    *2.8
%  7       33  33.4 23.9  -8.6   2.1     ?  ---- ---- -----  ----    MISS
%  8       48  33.8  1.5 -17.2   2.9    13  31.0  1.4 -19.6   5.8    *2.0
%  9       59  42.7 21.0 -26.5   3.8    16  36.5 21.7 -18.7   4.4    *1.2
% 10       80  59.6 23.0  -8.7  12       5 ===> another part of Sculptor
% 11      101  77.4  1.5 -18.1  29       1 ==> contained in 1
% 12      111 102.4 23.0  -7.9  90       3 ==> contained (?) in 3

A more recent work applying another void search algorithm, is that of 
\cite{lin95}. In this work single spheres, that are devoid of a certain type 
of galaxies (depending on the morphological type and luminosity), are used. 
The algorithm is applied to an area north of the super-galactic (SG) plane, 
showing that voids defined by bright elliptical galaxies have a mean diameter 
of up to 40\hmpc, in agreement with the \vf\ results. When considering 
fainter galaxies, the voids are smaller, with the faintest galaxies studied 
defining 8\hmpc\ voids, suggesting that faint galaxies delineate smaller 
voids within larger ones, which are defined by the bright galaxies.

%  ****************************************
%  *    Section 3: The \vf\ Algorithm     *
%  ****************************************

\section{The {\sc Void Finder} Algorithm}
\label{algo}

The \vf\ algorithm was designed with the following conceptual picture in mind: 
The main features of the LSS of the universe are voids surrounded by walls. 
The \emph{walls} are generally thin, \twod\ structures characterized by a high 
density of galaxies. They constitute boundaries between under-dense regions, 
generally ellipsoidal in shape---the \emph{voids}.
Although coherent over large scales, the walls---being subject to 
small-scale clustering---are not homogeneous and contain small breaches which
we wish to ignore. 
Galaxies within walls are hereafter labeled \emph{wall galaxies},
while the non-wall galaxies are named \emph{field galaxies}.
The voids are not totally empty: there are a few galaxies in them, which we
call \emph{void galaxies}.
The void galaxies are a sub-population of the field galaxies, indicating which 
of the field galaxies are in the voids. Wall galaxies, on the other hand, 
cannot---by definition---be in a void.

We define a void as \definition{a continuous volume that does not contain 
any wall galaxies and is nowhere thinner than a given diameter}. 
In other words, 
one can freely move a sphere with the minimal diameter all through the 
void. This definition does not pre-determine the shape of the void: it can 
be a sphere, an ellipsoid, or have a more complex shape, including
a non-convex one. The definition is targeted at identifying the same regions
that would be recognized as voids, when interpreting a point distribution
by eye. As the voids are defined based on the point distribution of 
galaxies, we do not need to introduce any smoothing scale. This is 
especially important since one of our goals is 
to measure the volumes of the voids. As the regions containing most of the 
galaxies---walls or filaments---are thinner than $ \sim 10 \hmpc $, even 
relatively small smoothing scales will spread the dense regions into the 
under-dense ones, artificially diminishing the voids.

Our voids may contain galaxies. A stiffer requirement, such that voids should 
be completely empty, is too restrictive as a single galaxy located in the 
middle of what we would like to recognize as a void might prevent its 
identification. However, for this definition to 
be practical we must be able to identify the field galaxies before we 
can start locating the voids.

The algorithm is divided into two steps. First the \wb\
identifies the wall galaxies and the field galaxies. Then the \vf\
locates the voids in the wall galaxy distribution. All
together, our method incorporates three parameters, which we
define below. Two of these parameters ($n$ and $\beta$) are used to
determine the field galaxies; the third parameter ($\xi$) is used during
the void search. Specific values for these parameters were chosen
after a trial-and-error procedure with various simulations (\sect{voro}), 
to give results that resemble as much as possible eye-estimates of the voids.

\subsection{The {\sc Wall Builder} Phase}
\label{wb}

We define a wall galaxy as \definition{a galaxy that has at least 
$n$ other wall galaxies within a sphere of radius $\ell$ around it}.
The radius $\ell$ is hereafter referred to as the
\emph{wall separation distance}.
A galaxy that does not satisfy this definition is classified as a field galaxy.
This is a recursive definition which we apply successively until all the 
galaxies are classified.
The minimal value that enables filtering of long thin chains of galaxies is 
$n = 3$\@. The choice $n = 2$ allows thin chains extending between dense 
structures. With $n = 3$ the minimal structure that can be recognized as a 
wall is a pyramid-like 4-galaxy cluster, with all distances between these 
galaxies smaller than $\ell$.
The wall separation distance $\ell$ is chosen in the following manner:
Let the distance to the $n$'th closest neighbor of a galaxy be $l_n$.
For a given sample this quantity has an average value $\bar{l}_n$ and a 
standard deviation $\sigma_n$. The radius $\ell$ is defined as:
\beeq  \ell \equiv \bar{l}_n + \beta \sigma_n  \eneq
We have chosen here $n = 3$ and $\beta = 1.5$. 

\begin{figure}
\centering
%graphicx: \includegraphics[width=\linewidth]{wb_fig.ps}
\plotone{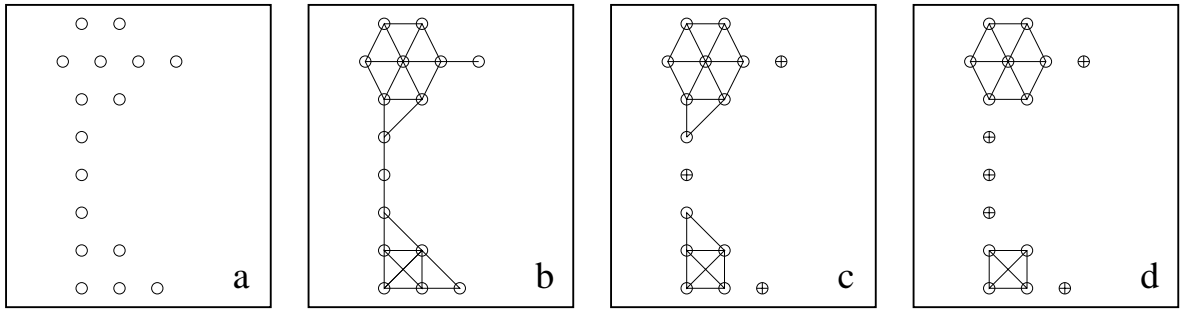}
\caption[Wall construction using the \wb]
{Wall construction using the \wb.
\emph{Panel~a}: A toy distribution of 16 galaxies ($\circ$).
\emph{Panel~b}: After the calculation of $\ell$, all galaxy pairs closer than 
this separation are marked.
\emph{Panel~c}: Galaxies with less than three neighbors are flagged as field 
galaxies ($\oplus$).
\emph{Panel~d}: The final result (after one additional iteration of neighbors 
count). The string extending between the dense structures has been eliminated.}
\label{wb-fig}
\end{figure}

\Fig{wb-fig} demonstrates how the \wb\ works for $n = 3$. 
Notice how the galaxy string is filtered, while the dense structures 
(upper and lower parts of each panel) are identified and maintained. 
Additional examples of the \wb, using Voronoi tessellations, are presented in 
\sect{voro}\@. As a side-bonus of this procedure, originally dedicated to 
filtering the field galaxies, we obtain a visual identification of the walls. 
This is done by drawing all the links between wall galaxies satisfying:
\beeq  \label{couples}
{\mathrm{dist}}(\vec x_i, \vec x_j) < \ell
\eneq
These connections are not used in the next step of void analysis, but provide 
us with another visual tool to examine our results (e.g., see 
\Fig{tess2-top-fig})\@.

\subsection{The {\sc Void Finder} Phase}
\label{vf}

The \vf\ searches for spheres that are devoid of any wall galaxies. 
These spheres are used as building blocks for the voids. 
A single void is composed of as many (or as few) superimposing 
spheres as required for covering all of its volume. The algorithm is
iterative, initially identifying voids containing relatively large
empty spheres, than proceeding to voids containing smaller
spheres. The process can be stopped at any desired \emph{void resolution},
defined as the diameter $d$ of the minimal sphere used as a void
building-block.
For practical reasons the actual process of improving the resolution is done 
in discrete iterations. We denote by \di\ the void resolution used during the 
$i$'th iteration. Thus, a void containing a sphere of diameter larger than 
\di\ would have been identified during the $i$'th iteration, if not earlier. 
On the other hand, if the largest sphere in a void is smaller than \di, that 
void would \emph{not} have been identified (yet).

The motivation for using this gradual process of refining the
resolution is the problem of keeping apart neighboring voids.
If voids were always well-separated by easily recognized
homogeneous over-dense regions, there would not be any reason
to go through this complicated process. One could simply start from
an initial sphere, located somewhere in the void, and work his way
from that sphere until he is stopped by a wall. Doing this in all
directions and from all spheres would surely be enough to encompass 
the whole void.
However, the situation is more complicated. The walls are 
not homogeneous, often missing a few bricks now and again. Our eyes are
capable of ignoring these breaches in the walls, still identifying
the voids as individual objects. For an algorithm aimed at objectively
identifying the voids, this is a major problem: how to keep
apart neighboring voids, when the boundary is not well marked?
A simple application of the above process 
would result in one all-connected void, since distinct voids would
merge with finger-shaped connections passing through the gaps
in the walls, in this way connecting those that should have been kept 
separate.

\begin{figure}
\centering
%graphicx: \includegraphics[height=0.3\linewidth]{triang1.ps}
%graphicx: \hfill
%graphicx: \includegraphics[height=0.3\linewidth]{triang2.ps}
\plottwo{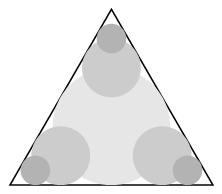}{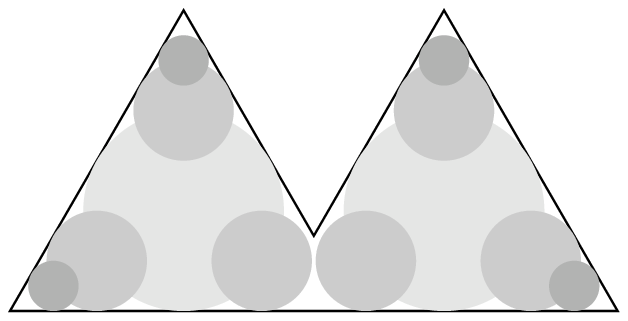}
\caption[Void coverage using spheres (triangular voids example)]
{Void coverage using spheres: Larger spheres (light gray) are identified 
first, and smaller ones (darker gray shades) are identified later. 
\emph{Left}: A triangular void. Seven spheres cover the void, the radius of 
the smallest spheres is a quarter of that of the largest one. 
\emph{Right}: Two triangles partially overlapping, as if there is a breach in 
the wall separating the voids. The gap is larger than the smallest sphere used 
to cover each void. However, we still can separate the two voids.}
\label{trig-fig}
\end{figure}

The size of the spheres per se is not a good criterion for differentiating
between a sphere which is a part of the void proper, and a sphere that is
passing through a wall making a false connection. To illustrate this we 
will use the somewhat artificial example of triangular voids (see 
\Fig{trig-fig}, left triangle).
To cover a large fraction of this void, one sphere is perhaps sufficient. 
But to reproduce faithfully the original void, several spheres with various 
diameters are required---some of the additional spheres being quite small, 
compared to the initial one.
Now, imagine a situation where the two voids are not ideally separated---%
the two triangles somewhat overlap (\Fig{trig-fig}, right-hand side). 
We find that spheres of the same size cover the volume of the voids near the 
remote edges, but also might connect the two voids to one.

By using a gradually refined void resolution, we 
overcome this problem in most situations. At the risk of being
tautological, we will note that the difference between the void proper
and a false connection is that the connections are characterized by
them \emph{being thinner than the rest of the void} and 
\emph{leading into another void}.
These two properties are critical for our
purpose. The first property allows for the initial identification of
the center of the void, before finding out about the connections.
Using a relatively crude resolution, we will first find the larger spheres 
that actually make up the void.
Now, when the resolution becomes sufficiently small to allow spheres to cross 
the walls, we will discover that the other side of the wall is already 
occupied by another void (identified earlier).

Hence, the spheres for covering a void are picked up in two stages: the 
\emph{identification stage}, followed by \emph{consecutive enhancements}.
We will now describe these stages in detail.\footnotemark

\footnotetext{This later addition of spheres to the central part of the 
voids is the main improvement here over the \vf\ algorithm that was presented 
earlier \citep{epd1}.
The older version, initially used for analyzing the SSRS2, did not include 
this feature. All the results presented here use the improved
version of the algorithm.}

The \emph{identification stage} identifies the central parts of the void.
Usually, these spheres cover only about half of the actual volume.
We focus (at this stage) on identifying a certain void as a separate 
entity, rather than trying to capture all of its volume.
The central parts of a void are covered using spheres with diameters
in the range $ \xi \dmax < d \leq \dmax $, with \dmax\ denoting the diameter
of the void's largest sphere. The parameter $\xi$ is the \emph{thinness 
parameter}. Once a group of such intersecting spheres has been dubbed a void, 
it will not be merged with any other group. The parameter $\xi$ controls the 
flexibility allowed at this stage. Setting $\xi = 1$ would leave
us with only the largest sphere in the void, while lowering $\xi$ allows
the addition of more spheres.
If the void is composed of more than one sphere 
(as is usually the case), then each sphere must intersect at least 
another one with a circle wider than the minimal diameter
$ \xi \dmax $. We have chosen $ \xi = 0.85 $, which allows for enough 
flexibility---without accepting counter-intuitive void shapes.
A lower $\xi$ reduces the total number of the voids, with a slow increase 
in their total volume.

After the central part of a void has been identified, we \emph{consecutively 
enhance} its volume, in order to cover as much of the void volume as 
possible using the current void resolution. These additional
spheres need not adhere to the $\xi$ thinness limitation: 
we scan the immediate surroundings of each void, and if empty
spheres are found then they are added to the void.
We scan for enhancing spheres of a certain diameter only \emph{after}
scanning for new voids with that diameter. In this way we do not falsely
break apart individual voids, and we do not prevent the identification
of truly new voids.
By applying consecutive enhancements to the voids, we have elaborated
our definition for a void, by allowing the voids to gradually become
thinner. Therefore the precise void definition is \definition{a continuous 
volume that does not contain any wall galaxies, and is thicker than an 
adjustable limit}. If this definition seems complex, it is a tribute
to the difficulty of teaching a machine how to grasp a composite
\tred\ object.

Using once more the triangular voids example (\Fig{trig-fig}, right hand 
side), notice that we do not falsely identify a new void near one of the 
edges of the triangle, since the initial sphere gradually takes-up 
that volume, leaving room only for smaller spheres which are well below the 
current void resolution.
Thus we identify only one void per triangle, as we should.
Further still, we do not connect the two voids, as they are
initially identified as two separate entities---not to be merged
later.

To conclude, we present a step-by-step listing of the various stages:
\begin{enumerate}
\item
Choose \di.
\item
Find all empty spheres with $ d > \xi \di $.
  \begin{enumerate}
  \item
  Construct a grid with a cell size $ (\xi \di / \sqrt{3})^3 $.
  \item
  Choose the centers $\{c_j\}$ of the empty grid cells that are not contained 
  within previously found voids.
  \item
  Starting from each $c_j$, construct a continuous path along which the radius 
  of the maximal empty sphere centered on the path continuously increases. The 
  path ends at a point $c_j^{\mathrm{max}}$ where the size of the 
  maximal sphere is a local maximum.
  \item
  Keep all maximal spheres having $ d > \xi \di $. The redundancy in the grid 
  size insures that every sphere with $ d > \xi \di $ contains at least one 
  $c_j$, so the identification of all such spheres is guaranteed.
  \end{enumerate}
\item
Arrange the spheres into groups (i.e., new voids), satisfying:
  \begin{enumerate}
  \item Each group contains at least one sphere with $ d \geq \di $, the 
        largest of which is denoted \dmax.
  \item A group may include additional spheres; these intersect at least one 
        other sphere in the group with a circle whose diameter is larger 
        than or equal to $ \xi \dmax $.
  \end{enumerate}
\item
Enhance the older voids:
  \begin{enumerate}
  \item Scan the outskirts of all previously identified voids for empty 
        spheres.
  \item Add $ d > \di $ spheres that intersect a void to it.
  \end{enumerate}
\item
Decrease the void resolution \di, and iterate the process.
\end{enumerate}

\begin{figure*}
\centering
%graphicx: \includegraphics[width=0.975\linewidth]{demo.ps}
\plotone{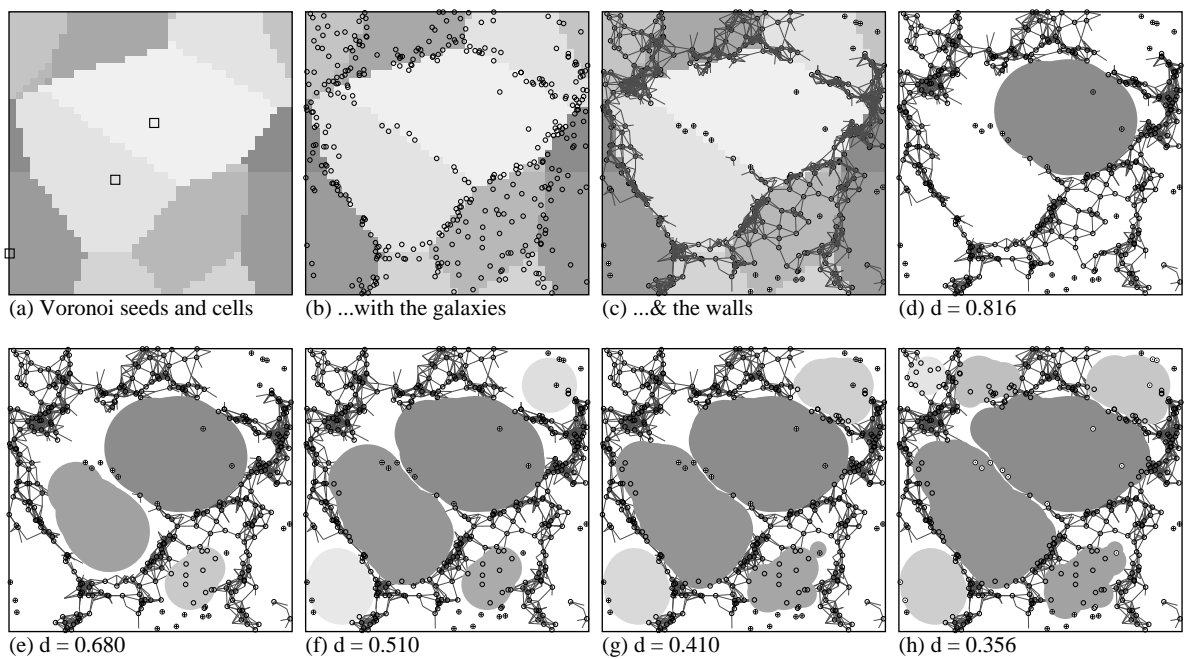}
\caption[Void coverage with the \vf]
{A demonstration of the way the \vf\ covers the voids. All panels depict the 
same slice, cut through a certain Voronoi tessellation. We present the Voronoi 
seeds ($\Box$) \& cells (panel~a), the galaxies (panel~b) and the walls 
(panel~c). The remaining panels show the voids' image, at various void 
resolutions \di. More voids are recognized as we refine \di, and the older 
voids are enlarged.}
\label{vf-demo-fig}
\end{figure*}

\Fig{vf-demo-fig} demonstrates how the \vf\ works. In all panels
we show the same slice through a \tred\ Voronoi Tessellation (see 
\sect{voro} for details), and we follow the development of the
void image as we refine the void resolution. Initially, we locate the
voids containing the largest empty spheres. In the following iterations we 
locate the smaller voids, and---when appropriate---enlarge the volumes of the 
older ones.

One must take into account the boundaries of the distribution. We treat the 
limits of the distribution as rigid boundaries. This may cause some distortion 
in the voids found close to the boundaries (causing them to be smaller than 
the rest). Still, this is the least speculative option as we do not make any 
assumptions about unsurveyed regions.

\subsection{Statistical Significance}
\label{stat}

To assess the statistical significance of the voids we compare the voids found 
in observed data with voids found in equivalent random distributions.
The random distributions mimic the sample's geometry and 
density, and are analyzed by the algorithm in exactly the same manner.
Averaging over the random catalogs we calculate \npoisson, the expected number 
of voids in a Poisson distribution as a function of the void resolution $d$.
We compare this with the observed number, \nsurvey\@.
We define the \emph{confidence level} as:
\beeq \label{confid}
p(d) = 1 - \frac{\npoisson}{\nsurvey}
\eneq
The closer $p(d)$ is to unity, the less likely the void 
could appear in a random distribution.
We consider voids with $ p \geq 0.95 $ as statistically significant.
At a certain void resolution \dstop, \npoisson\ exceeds \nsurvey,
and we terminate the void search.

The statistical significance defined in this way is attributed according to the
\emph{order} in which the voids were identified, not according to their
sizes.
This is so because the parameter determining how early a void is detected is 
the diameter of the largest sphere it contains, and not its
total volume. Of course the total volume of a void and the diameter of the 
largest sphere contained in it are correlated. However, a void composed of a 
single large sphere will be 
detected earlier---and hence considered more significant---than a 
\emph{larger} void composed of several spheres all having smaller radii.
This is in agreement with the fact that clear spherical voids are more 
prominent when a galaxy distribution is inspected by eye. In retrospect
this choice is also justified by the theoretical expectation that voids 
become more spherical with cosmological time \citep{blu92}.

Voids in the actual surveys are systematically larger than the voids in the 
random distributions. When we reach the void resolution \dstop\ (where the 
\emph{number} of voids in the random distributions exceeds the void number in 
the actual survey), the \emph{volume} taken up by the random voids is still 
much smaller than the volume of the true voids. By ignoring this fact we 
actually under estimate the significance of the voids.

\subsection{Redshift Surveys}
\label{surveys}

The average galaxy number-density decreases with depth in a magnitude-limited 
redshift survey. If not corrected, this selection effect will interfere with 
the algorithm in the deeper regions of the sample: field galaxies will occur 
more frequently, and the derived size of the voids will be larger. 
Consequently, systematically larger voids will be found at greater distances.

To avoid these effects, one should use a volume-limited sample, in which 
the galaxy number-density is constant and independent of the distance. A 
volume-limited sample with $ M \leq M_o $ has a depth:
\beeq 
\rvol = \sqrt{\lum_o / 4 \pi \flux_{m_b}} = 10^{-5-0.2(M_o-m_b)} \, \mbox{Mpc}
\eneq
where $m_b$ is the survey's magnitude limit and $\lum_o$ is the luminosity 
that corresponds to $M_o$.
However, current volume-limited samples are too small to study the LSS. For 
example, the SSRS2 sample contains 3162 galaxies with $ r < 130 \hmpc $. A 
volume-limited sample based on this distribution with such a depth 
($ M_o = -20.1 $) retains only 528 galaxies.
To overcome this, we use a semi--volume-limited sample: volume-limited up to 
some medium radius \rvol, and magnitude-limited beyond. We choose the depth 
\rvol\ by maximizing the number of bright galaxies $N(M \leq M_o)$:
\beeq  
%N(M \leq M_o) = \frac{4 \pi}{3} \rvol^3  \cdot  \eta \, \Gamma(1-\alpha, x_o)
N(M \leq M_o) = \frac{4 \pi}{3} \rvol^3  \cdot  \eta \Gamma(1-\alpha, x_o)
\eneq
where $\eta$ is the galaxy number-density. The incomplete $\Gamma$-function 
arises from the integration of the appropriate Schechter function, with 
$ x = \lum / \lum_\star $.

\begin{figure}
\centering
%graphicx: \includegraphics[width=0.95\linewidth,clip]{n_select.new.eps}
\plotone{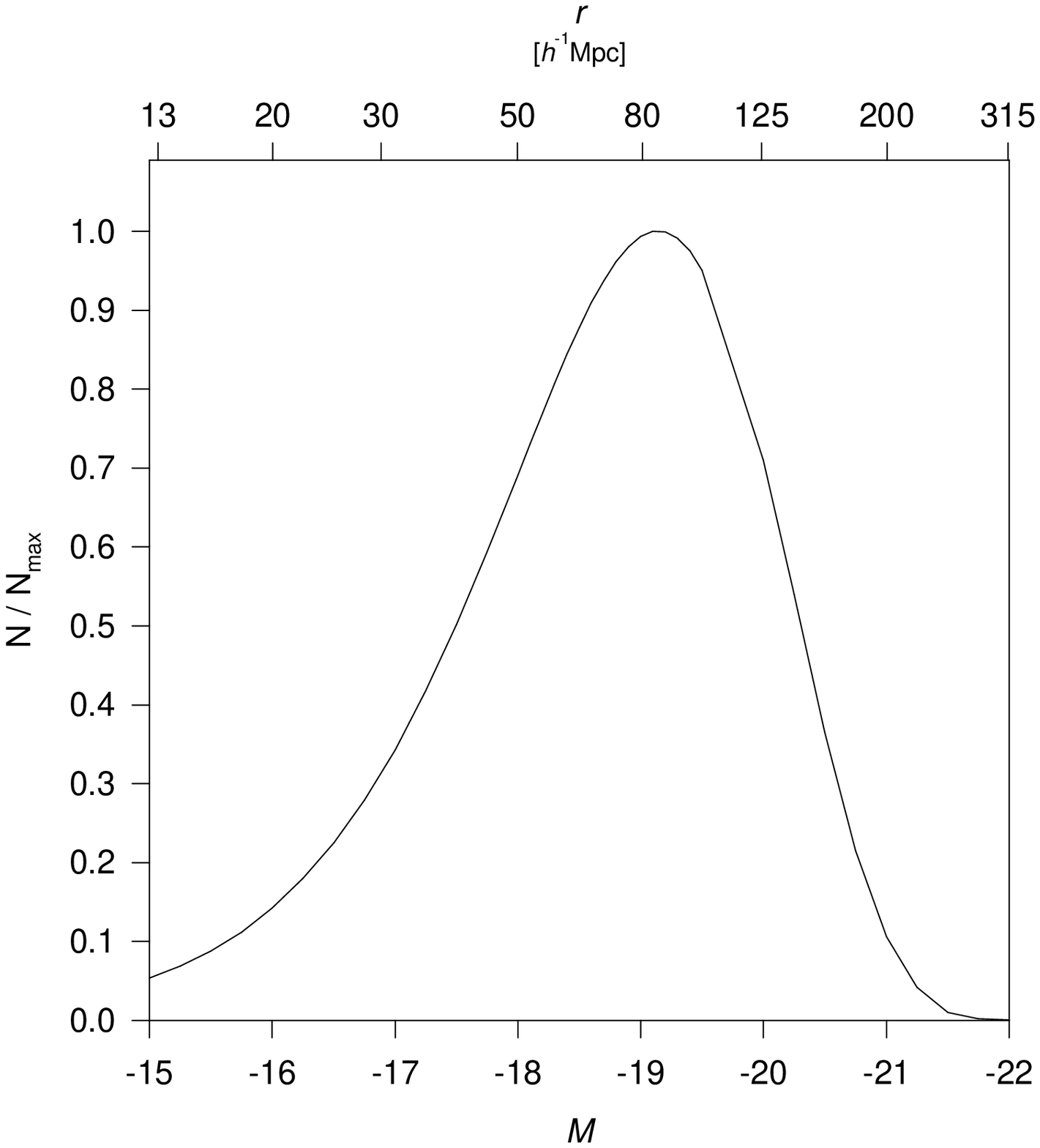}
\caption[Setting the depth of the volume-limited region]
{The relative number $ N / N_{\mathrm{max}} $ in the SSRS2 sample 
($ m_b = 15.5 $, $ \alpha = 1.2 $ and $ M_\star = -19.5 $) 
as a function of the depth of the volume-limited region.
The number of galaxies $N$ peaks at $ M = -19.1 $ ($ r = 83 \hmpc $).}
\label{l-func-fig}
\end{figure}

In \Fig{l-func-fig} we plot the normalized $N(M_o)$ for the SSRS2 parameters: 
$ \alpha = 1.2 $, $ M_\star = -19.5 $ and $ m_b = 15.5 $. In this example $N$ 
peaks close to $M_\star$ at $ M = -19.1 $, which corresponds to $x_o = 0.69$ 
and a depth $ \rvol = 83 \hmpc $.
If we increase the depth of the volume-limited sample, the number of 
galaxies decreases.

\begin{figure}
\centering
%graphicx: \includegraphics[width=\linewidth,clip]{ssrs2slf.eps}
\epsfig{file=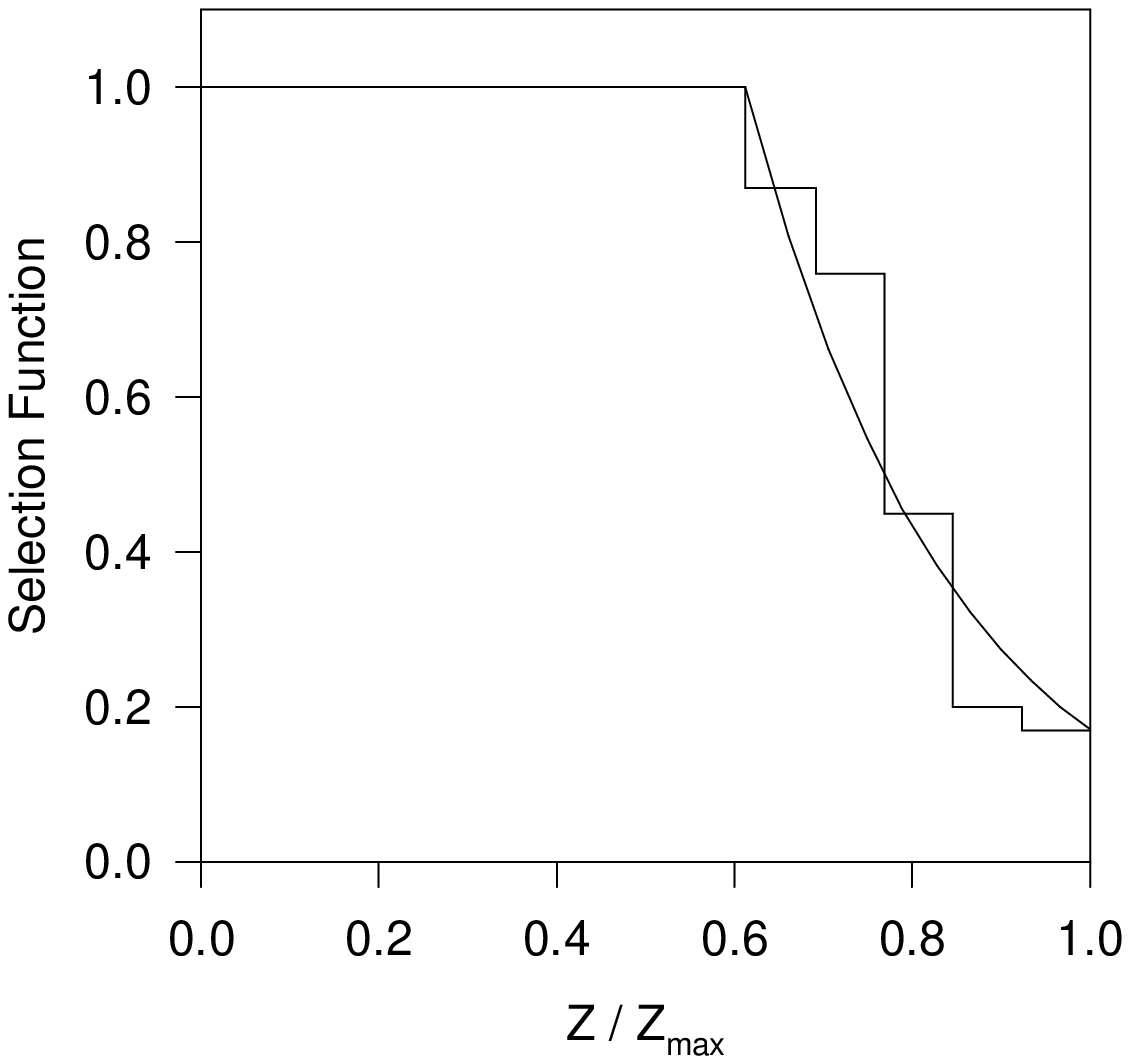, clip=}
\caption[SSRS2 selection-function]
{The SSRS2 selection-function, with $ \rvol = 79.5 \hmpc $ and 
$ \rmax = 130 \hmpc $. The smooth line depicts $\phi$, where as the dashed
line is the actual relative density, evaluated at 10\hmpc\ intervals.}
\label{ssrs2-sf-fig}
\end{figure}

No corrections are needed in the volume-limited region. We determine the 
values for $\ell$ and \di\ in this region. Beyond \rvol\ we define $\phi(r)$, 
a selection-function based on the Schechter luminosity-function: 
\beeq  \phi(r) = \frac{ \Gamma(1-\alpha, x_M) }
                      { \Gamma(1-\alpha, x_{M_o}) }  \eneq
where $ x_M = 10^{-0.4(M - M_\star)} $.
The selection-function $\phi(r)$ is the observed fraction of galaxies at the 
distance $r$, relative to \rvol. A plot of the selection-function for the 
semi--volume-limited sample for SSRS2 is shown in \Fig{ssrs2-sf-fig}.
Using the selection-function, we modify both phases of the algorithm. In the 
\wb\ phase, we consider larger spheres when counting neighbors at 
$ r > \rvol $. The radius of the counting-spheres is modified:
\beeq  \ell' = \left\{ \begin{array}{ll}
                          \ell                 & \mbox{if $ r < \rvol $} \\
                          \ell / \phi^{1/3}(r) & \mbox{otherwise}
                       \end{array}
               \right.              \eneq

A similar correction is applied to the \vf\ phase. A void of a given size 
found in a low density environment is less significant than a void of the 
same size found in a high density environment. In order that all the voids 
found in a given iteration are equally significant, we adjust the algorithm 
so that at a given iteration relatively larger voids are accepted, if 
located at $ r > \rvol $:
\beeq  \label{scale-eq}
       d'_{\mathtt{i}} = \left\{
                \begin{array}{ll}
                   d_{\mathtt{i}}                 & \mbox{if $r < \rvol$} \\
                   d_{\mathtt{i}} / \phi^{1/3}(r) & \mbox{otherwise}
                \end{array}
                          \right.
\eneq

\begin{figure}
\centering
%graphicx: \includegraphics[width=\linewidth,clip]{ssrs2rnd.eps}
\epsfig{file=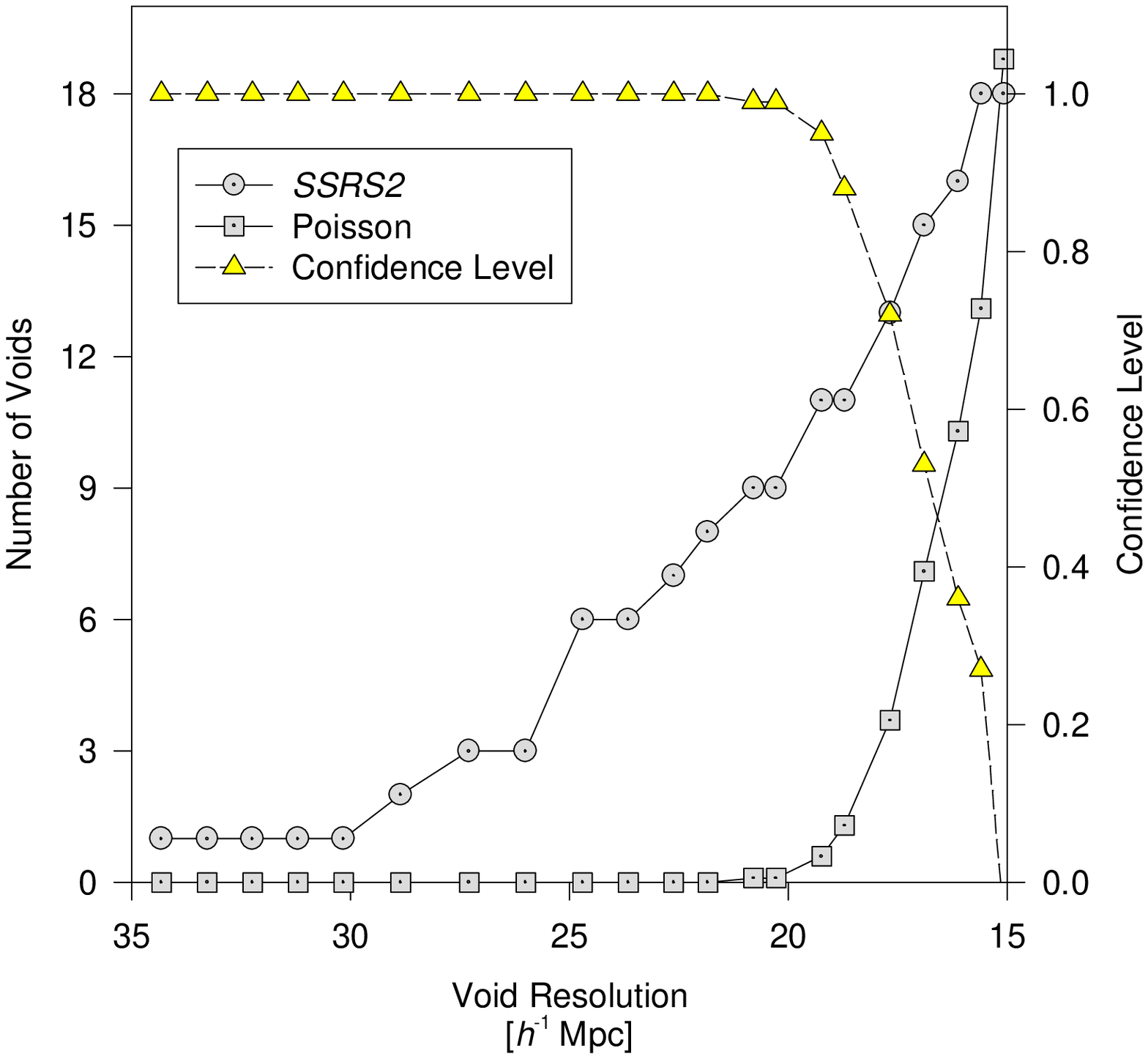, clip=}
\caption[SSRS2 voids confidence level]
{The accumulated number of voids as a function of the void resolution 
$d$, for the SSRS2 and for equivalent random catalogs. The derived 
confidence level $p$ is also indicated.}
\label{ssrs2-rnd-fig}
\end{figure}

As mentioned earlier (\sect{stat}), we use random distributions that mimic 
the true sample's geometry and density, in order to assess the statistical 
significance of the voids. This scheme needs to be fitted, to correspond to 
the decrease in the galaxy density with $r$. We continue to use the same 
definition for the confidence level $p$ (\eq{confid}). However, $d$ is now a 
function of $r$, scaled in the same manner given by \equ{scale-eq} (see 
\Fig{ssrs2-rnd-fig}).

%  ****************************************
%  *   Section 4: Voronoi Distributions   *
%  ****************************************

\section{Voronoi Distributions}
\label{voro}

As a test-bed for the \vf\ algorithm, we use Voronoi distributions: A
distribution of galaxies that is based on a Voronoi tessellation
\citep{vor08}.
A Voronoi tessellation is a tiling of space into convex polyhedral 
cells, generated by a distribution of seeds. 
To generate a galaxy distribution in which the galaxies are located on the 
walls of the Voronoi cells, we have used an algorithm developed by 
\cite{vdw89}. The resultant galaxy distribution has the desired 
characteristic of large empty regions (i.e., voids), which we identify by 
the \vf\ algorithm.

\begin{figure*}
\centering
%graphicx: \includegraphics[width=0.913\linewidth]{tes2_bas.ps}
\epsfig{file=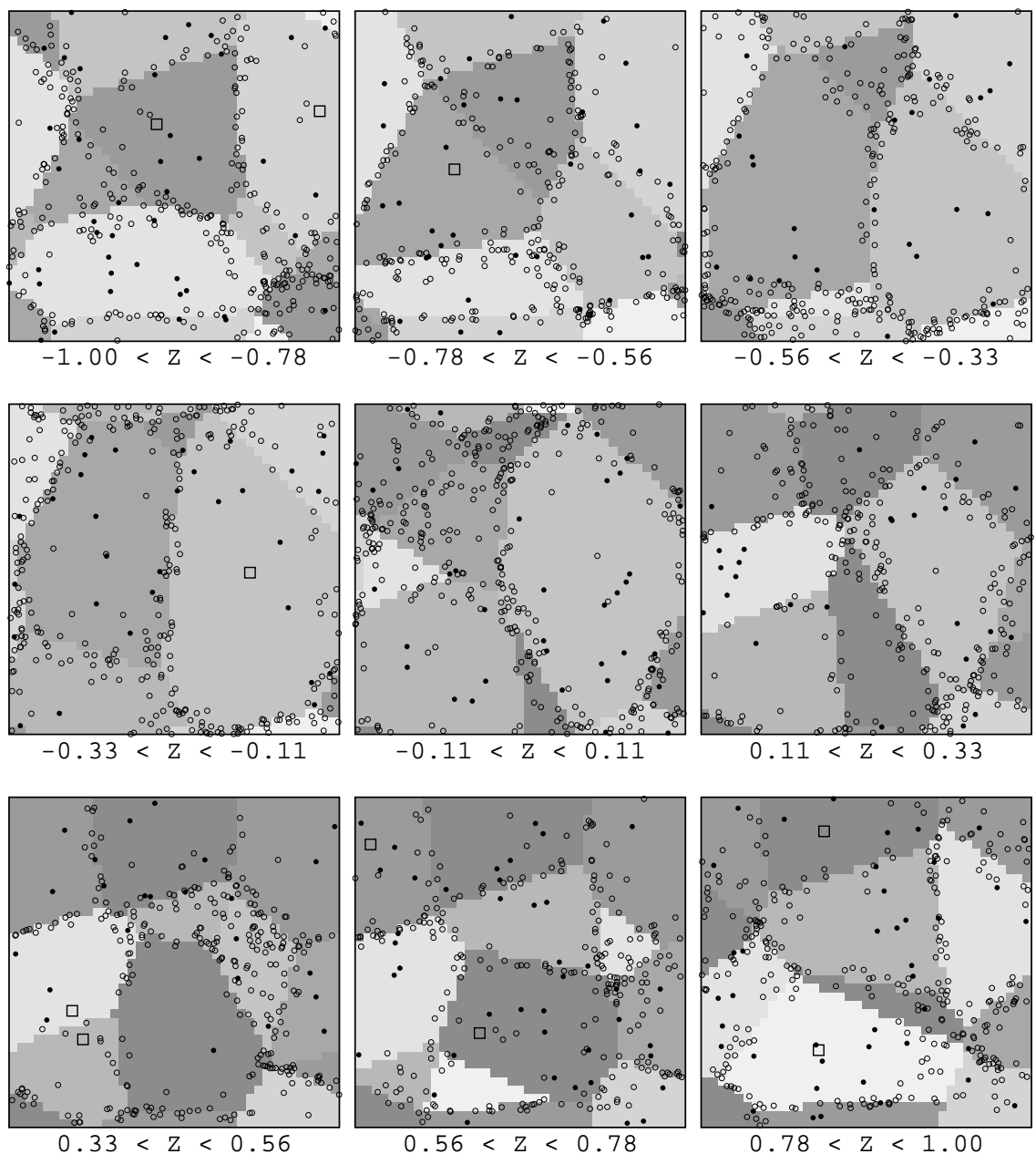, width=0.913\linewidth}
\caption[A cubic Voronoi tessellation: Voronoi cells and galaxies]
{Nine planar slices showing Voronoi cells and galaxies in cubic Voronoi 
tessellation, generated from 10 seeds with 3000 galaxies (10\% random). The 
cells are depicted using various shades of gray, indicating the intersection 
of the Voronoi cell with the plane at the center of the slab. The locations of 
the Voronoi seeds are marked by `$\Box$'. Galaxies associated with cell 
boundaries are marked by `$\circ$', and random galaxies by `\textbullet'.}
\label{tess2-base-fig}
\end{figure*}

A \emph{Voronoi tessellation} is constructed from a given set of 
\emph{seeds} $ \{ \vec x_i \} $.
Based on the locations of these seeds, we divide the volume into cells.
The Voronoi cell $\Pi_i$ of seed $i$ is defined by the following set of
points $\vec x$:
\beeq  \Pi_i = \{ \: \vec x \;\;\: | \;\;\: 
                  {\mathrm{dist}}(\vec x, \vec x_i) <
                  {\mathrm{dist}}(\vec x, \vec x_j) \quad \;
                  \mbox{for all $ j \neq i $} \: \}         \eneq
%In other words, $\Pi_i$ is the set of points that is nearer to the seed
%$\vec x_i$ than to any other seed (the cells thus created are also known 
%as Wigner-Seitz cells, among other names).
In other words, $\Pi_i$ is the set of points that is nearer to the seed
$\vec x_i$ than to any other seed.
In \Fig{tess2-base-fig} we show \twod\ cuts through a Voronoi tessellation.

We assign a finite width to the walls and position the galaxies on 
the boundaries between the Voronoi cells, with a Gaussian displacement 
in the distance from the exact cell boundary. 
These galaxies are designated to be \emph{wall galaxies}, but do not
necessarily end-up in this way: in regions where the galaxy distribution
is sparse, the \wb\ may identify these galaxies as field galaxies.
Further still, if the galaxy density is low, there may be boundaries between 
Voronoi cells with no galaxies at all.
Additional random galaxies correspond to field galaxies.
Most of these randomly placed galaxies end-up as field galaxies. A few, 
located in dense areas, will be identified by the \wb\ as wall galaxies.

We will call a galaxy distribution constructed in this way a \emph{Voronoi 
distribution}. The location and number of 
the Voronoi cells (the would-be voids), the spread of the wall galaxies 
and the fraction of random galaxies are all known. Therefore, we 
can use this distribution as a test bed for our algorithm.

\begin{figure*}
\centering
%graphicx: \includegraphics[width=0.913\linewidth]{tes2_top.ps}
\epsfig{file=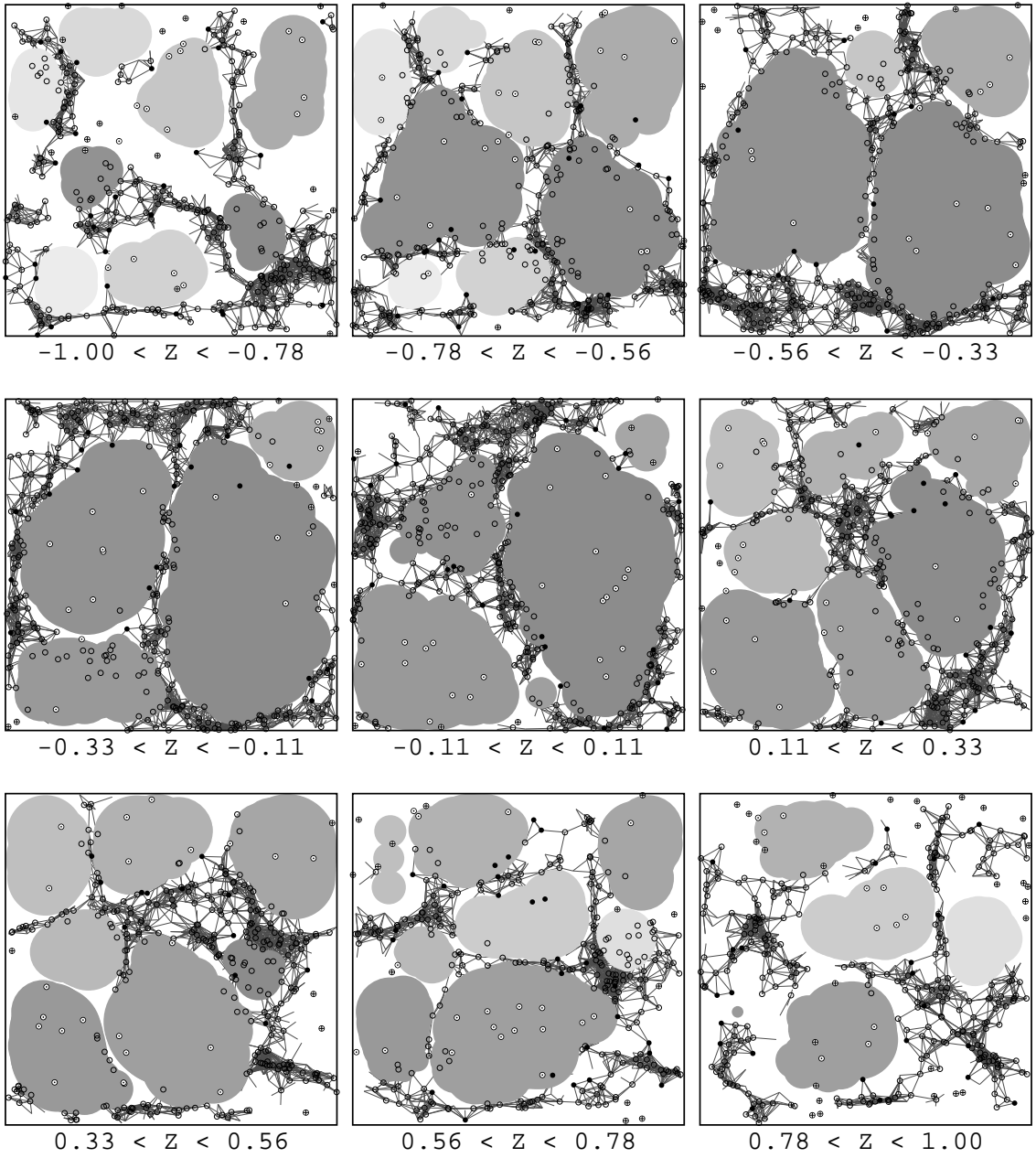, width=0.913\linewidth}
\caption[A cubic Voronoi tessellation: The reconstructed voids]
{The \vf\ reconstruction corresponding to \Fig{tess2-base-fig}:
The voids are indicated using various gray shades, where the depicted 
voids correspond to the intersection of the central plane of each slab
with the \tred\ voids.
Also shown are the walls (dark lines---connections between nearby galaxies 
which satisfy \eq{couples}) and the galaxy breakdown 
(wall galaxies: `$\circ$' or `\textbullet'; field galaxies, outside the 
voids: `$\oplus$'; void galaxies: `$\odot$').
Non-void galaxies that look as if they are in the voids are actually 
\emph{not}---they only seem so, due to projection effects.
Voids in the first and last panels are smaller due to boundary effects.}
\label{tess2-top-fig}
\end{figure*}

We have constructed a Voronoi distribution using 3000 galaxies and 10 seeds, 
which includes 300 (10\%) random galaxies. These numbers correspond roughly 
to what is available in present redshift surveys.
The original Voronoi tessellation (\Fig{tess2-base-fig}) is compared to the 
\vf\ reconstruction (\Fig{tess2-top-fig}).
All Voronoi cells are reproduced except the very small cells near the
boundaries, that were cut by the box limit.
The reconstructed voids follow closely the original Voronoi cells, 
withstanding the noise introduced by the random galaxies. 
The walls highlighted using the \wb\ are located along the boundaries between 
the Voronoi cells. 

It is worthwhile to examine where the algorithm fails. A single void was 
broken into two (\Fig{tess2-top-fig}, lower-left corner of upper-left box)
due to some random galaxies that extended the walls into this cell.
Also note the instances where two Voronoi seeds are situated relatively close 
to each other (e.g., in the lower-left box). In such a situation the resultant 
wall is very sparse. This is due to the way in which the Voronoi 
code generates the galaxy distribution.

\begin{figure}
\centering
%graphicx: \includegraphics[width=\linewidth]{vorosurv.ps}
\plotone{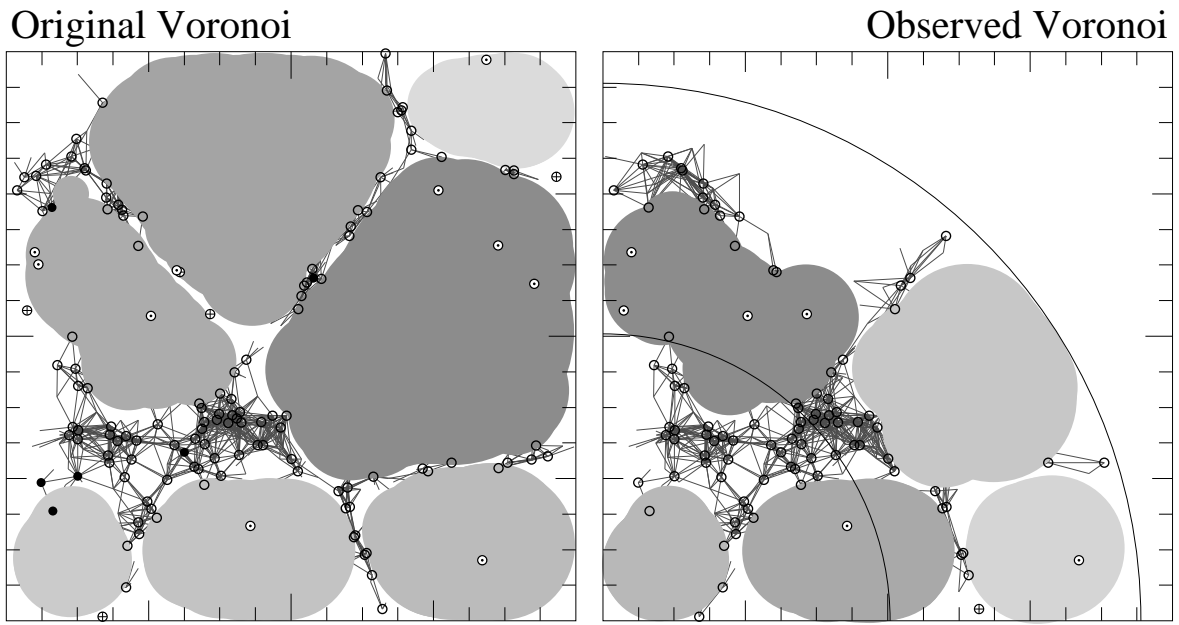}
\caption[Voronoi-based mock distribution]
{\emph{Left}: \vf\ reconstruction of an ordinary Voronoi distribution.
\emph{Right}: reconstruction of a mock survey based on the same data. The
intermediate arc marks the volume-limited region of the mock sample.}
\label{voro-surv}
\end{figure}

We have also created mock surveys, based on Voronoi distributions. Galaxies 
in the Voronoi distribution were assigned magnitudes according to a Schechter 
function. Then, a magnitude-limited sample was chosen. To this Voronoi-based 
mock survey we have applied our usual procedure: analyzing a semi--volume-%
limited sample, applying corrections beyond \rvol. \Fig{voro-surv} shows a 
reconstruction based on the original data, along with a reconstruction based 
on the corresponding mock survey. The fit between the Voronoi cells and the 
recovered voids is still good, showing the adequacy of our method in 
analyzing actual surveys.

All together, the Voronoi tessellations that we examined show that the \vf\ 
indeed generates a faithful reproduction of the Voronoi cells. Further still, 
in cases where the reproduction merges adjacent Voronoi cells into one void, 
we see this as the adequate outcome of a missing wall. If we would have 
examined such a galaxy distribution by eye, with no prior knowledge about the 
locations of the Voronoi cells, we too would most likely consider that 
volume---originally occupied by two Voronoi cells---as one void. A level of 
$\sim 10\%$ random galaxies is tolerated, with no significant distortion in 
the void reproduction, and Voronoi-based mock surveys are also reproduced 
faithfully.

Boundary distortions are the cause for most of the cases in which the \vf\ 
departs from the original tessellation. This is also evident when considering
the void volumes:
The volume occupied by the voids is $ \sim 15\% $ larger, if we consider only
an inner cube and not the full tessellation. Voids near the boundaries are
typically smaller (if at all recognized), as the Voronoi cells are bisected
by the cube's boundary.

%  *****************************************
%  *     Section 5: Voids in the SSRS2     *
%  *****************************************

\section{Voids in the SSRS2}
\label{ssrs2}

The SSRS2 survey \citep{dc94} consists of $\sim 3600$ galaxies with 
$ m_b \leq 15.5 $ in the region $ -40 \arcdeg < \delta < -2\fdg5 $ and 
$ b \leq -40 \arcdeg $, covering $ 1.13\, \mbox{sr} $\@. We have considered a 
semi--volume-limited sample, in this case consisting of galaxies brighter 
than $ M_o \leq -19 $, corresponding to a depth $ \rvol = 79.5 \hmpc $. The 
Schechter luminosity function was evaluated with $ M_\star = -19.5 $ and 
$ \alpha = 1.2 $, as derived for the SSRS2. Our final semi--volume-limited 
sample consists of 1898 galaxies, extending out to $ \rmax = 130 \hmpc $ 
where the selection-function $\phi$ has dropped to 17\% (\Fig{ssrs2-sf-fig}). 
It should be emphasized that the SSRS2 analysis is performed in 
\emph{redshift-space}. However, because of the paucity of large clusters and 
the small amplitude of peculiar motions in the volume surveyed by the SSRS2, 
redshift distortions are small \citep{dc97} and the properties derived here 
should reflect those of voids in real-space.

The \wb\ analysis of the SSRS2 has classified 1736 galaxies (91.5\%) as wall 
galaxies, and 162 (8.5\%) as field galaxies. The wall separation distance was 
$ \ell = 7.4 \hmpc $. In the volume-limited region we have 
$ n^{-1/3} = 6.4 \hmpc $, so $ \ell / n^{-1/3} = 1.16 $. The wall galaxies are 
grouped in ten structures. The remarkable fact that one structure contains 
most (96\%) of the wall galaxies will be discussed elsewhere, as this may have 
interesting statistical implications regarding the connectivity of the 
wall--filament skeleton. The rest of the wall galaxies are found in nine 
groups, each having 4 to 21 galaxies.

We have identified eleven significant ($ p \geq 0.95 $) voids within the 
volume probed by the SSRS2\@. These voids were detected while the void 
resolution was $ d \geq 19.2 \hmpc $. In the following calculations we 
take into account only these voids, unless otherwise specified. Seven 
additional voids were identified before the void search was terminated at the 
resolution $ \dstop = 15.1 \hmpc $, for which $p$ vanishes. 
\Fig{ssrs2-rnd-fig} plots the accumulated number of voids in the SSRS2 and in 
the corresponding random catalogs, as a function of the void resolution 
$d$. The locations and characteristics of all eighteen voids are given in 
Table~\ref{ssrs2-table}. Column (1) lists the confidence level $p$ attributed 
to each void. The diameters given in column (2) are of a sphere with the same 
volume as the whole void, as is listed in column (3). The center of the void 
given in columns (4)--(6) is defined as its center-of-(no)-mass. The density 
contrast value listed in column (7) was corrected for the average galaxy 
density at the same distance as the center of the void. In column 
(8) we give the fraction of the total volume of the void covered by the 
single largest sphere contained in it. This value is typically $ \sim 40\% $
of the total volume of the void. Finally, in column (9) we indicate the 
corresponding voids identified in \citeauthor{dc88} (\citeyear{dc88}, 
hereafter dC88).

The average size of the voids in the survey as estimated from the 
equivalent diameters is $ \bar{d} = 40 \pm 12 \hmpc $.
The average under-density within the voids was 
found to be $ \delta \rho / \rho \approx -0.9 $, a quite remarkable result 
showing how empty voids are of bright galaxies.
The eleven significant voids comprise 54\% 
of the survey's volume. An additional 5\% is covered by the seven
additional voids, totaling in $ \sim 60\% $ of the volume being occupied
by these voids.
We estimate that the walls occupy less than 25\% of the volume.

The SSRS2 sample and the resultant \vf\ reconstruction is presented in 
\Fig{ssrs2-fig}, as six $ 6\fdg25 $-wide constant-declination slices.
Do not be mislead by the wide right-ascension range of this survey, 
emphasized by these images. In declination the survey spans $ 37\fdg5$ 
(see \Fig{ssrs2-ra-fig}), limiting the sizes of the voids we find.
The slices contain numerous walls and 
filaments. The nearest prominent features, the Fornax cluster and the 
Eridanus group, can easily be recognized at ($ r \sim 10 \hmpc, \alpha = 
3\fh5 $) running through most of the declination range. One of the 
outstanding structures in this survey is the Southern Wall (SW). This wall 
(dC88) is the dense structure running across all slices, 
from ($ \alpha = 4\hour $, $ r = 45 \hmpc $) to ($ \alpha = 0\hour $, $ r = 
90 \hmpc $). It is especially prominent in the $ \delta > -15\arcdeg $ and 
the $ \delta < -27\fdg5 $ slices. The Pavo-Indus-Telescopium (PIT) 
supercluster runs along the line of sight at $ \alpha = 21\fh5 $ in the 
$ \delta < -21\fdg3 $ slices. These two walls are almost parallel, and are 
connected through a third wall running perpendicular to the line of sight at 
$ r \sim 90 \hmpc $ in the right ascension range $21\fh5$ to $0\hour$. These 
walls bound two voids (void~5 and void~12 in Table~\ref{ssrs2-table}), 
identified earlier as one void (void~3 in dC88)---here the void is broken 
into two by a filament almost connecting PIT and the SW. A third structure 
runs parallel to the SW, bounding another void (void~1) between them.

The largest void found in the SSRS2 survey (void~3) has an equivalent 
diameter $ d = 60.8 \hmpc $, making it comparable in volume to the large void 
found in the \bootes\ \citep{kir81}. This void is an ellipsoid, whose major 
axis is perpendicular to the line of sight, located at: 
$ 80 \hmpc < r < 130 \hmpc $;
$ -25\arcdeg < \delta < -2\fdg5 $; 
$ 21\hour < \alpha < 23\fh75 $. 
This void might actually be larger, since it is bounded (in three directions) 
by the limits of the SSRS2. A second large void (void~2, with $ d = 56.2 
\hmpc $) is also comparable to the \bootes\ void.

When preparing the semi--volume-limited sample, we cast off all faint 
$ M > M_o $ galaxies in the region $ r < \rvol $. These galaxies comprise the 
bulk of the surveyed galaxy population that we are forced to ignore (the rest 
are $ r > \rmax $ galaxies, where the sample is too sparse). Although we 
cannot use these galaxies during the analysis phases, we can still try and 
benefit from them \emph{a~posteriori}: after the voids are located, we 
examine the locations of these galaxies. In this way we have divided the 
original galaxy distribution in the volume-limited region into two distinct 
populations, according to some absolute-brightness limit. 

There are 1264 $ M_o > -19 $ galaxies in the volume-limited region of the 
SSRS2. Almost 61\% of this region is covered by voids---but only 19\% of the 
faint galaxies are found within them. Even though the \vf\ algorithm uses 
only the brighter galaxies in this region, we find that the faint galaxies do 
not fill the voids, providing an excellent verification of our algorithm.
Still, the percentage of faint galaxies within the voids is significantly 
larger than that of the bright galaxies: only $ \sim 5\% $ of the bright 
$ M_o \leq -19 $ galaxies are contained in the voids.
% in detail: 241 faint galaxies in voids, out of 1264 (19%)
%            bright's: r<r_0:  38/ 783 = 4.9%
%                      all  : 118/1898 = 6.2%

%  ****************************************
%  *        Section 6: Discussion         *
%  ****************************************

\section{Discussion}
\label{discuss}

We have used the \vf\ algorithm to analyze two redshift surveys: The SSRS2 
and the \iras\ 1.2~Jy survey \citep{epd2}. These surveys represent two 
extreme cases in the trade-off between density and sky coverage. The SSRS2 is 
densely sampled ($ m_b \leq 15.5 $), but it has narrow angular limits, 
especially in the declination range. As a result, voids are often limited by 
the survey's boundary, diminishing their scale. The \iras\ is an almost 
full-sky survey ($87.6\%$ coverage), but it is rather sparse. As a result one 
cannot use a small void resolution, for lack of statistical significance. In 
addition to the above differences, the SSRS2 galaxies are optically selected, 
as opposed to the \iras\ galaxies.

Withstanding these differences, the results obtained with these surveys are 
similar. First, the surveys agree regarding individual voids in the regions 
where the surveys overlap. \Fig{iras-ssrs2-fig} depicts the redshift-space 
voids in the SG plane, for the \iras\ and for the corresponding part of the 
SSRS2\@. In the region where the SSRS2 sample overlaps the \iras\ sample, we 
find three of the eleven significant voids identified in the SSRS2. The 
corresponding \iras\ voids are $ \sim 33\% $ larger the SSRS2 ones, since 
they are not bounded by narrow angular limits as the SSRS2 voids.

Additionally, the results agree viz a viz the statistical characteristics of 
the voids:
\begin{enumerate}
\item Large voids occupy $ \sim 50\% $ of the volume.
\item Walls occupy less than $ \sim 25\% $ of the volume.
\item A void scale of at least 40\hmpc, with an average under-density of 
      $-0.9$.
\item Faint galaxies do not ``fill the voids'', but they do populate them 
      more than bright ones. 
\end{enumerate}
The void scale derived in both surveys is a lower limit: for the SSRS2, 
because of the narrow boundaries limiting the voids; and for the IRAS, due to 
the larger \dstop, and because of the conservative analysis applied regarding 
the ZOA and the field galaxies. Further still, regarding the walls, both 
surveys poses the remarkable characteristic where almost all ($ \sim 95\% $)
of the wall galaxies are contained in a single structure.

The fact that both the \iras\ and the SSRS2 are consistent regarding the void 
statistics as well as the individual voids is not trivial, since the \iras\ 
galaxies represent a special galaxy class, which is biased relative to the 
optical galaxies \citep{la88,lnp90}. The agreement between the surveys 
suggests that a similar void scale exists for both optically and \iras\ 
selected galaxies.

The \iras\ data has been used to derive the smooth density field, and it 
probes a volume comparable to that used to determine the density field of the 
underlying mass distribution from the \potent\ reconstruction method 
\citep{de90} based on the measured galaxy peculiar velocity field. The voids 
and walls identified by our algorithm \citep{epd2} indeed correspond to the 
under-dense and over-dense regions in the \iras\ density field \citep{sw95} 
respectively. Comparison with the SFI sample \citep{dc96} also demonstrates 
that the voids delineated by galaxies correspond remarkably well with the 
under-dense regions in the reconstructed mass density field derived from 
peculiar velocity measurements (but also compare with the Mark~III 
map---\citealt{de94,de97}).

Most of the over-dense regions, walls and filaments, are narrower than 
10\hmpc. The smoothing scale used for creating the density fields spreads the 
originally thin structures over wider regions, extending into the under-dense 
volumes. This has the effect of giving a false impression of a rather blurred 
galaxy distribution, where prominent over-dense structures are separated by 
small under-dense regions. The true picture is very different: there is a 
sharp contrast between the thin over-dense structures which occupy only the 
lesser part of the volume, and the large voids. The notion of a void filled 
universe cannot be avoided in this picture.

We have developed and tested a new tool for quantifying the large-scale 
structure of the universe. We focus on the under-dense regions, and for the 
first time we have individually identified and statistically quantified the 
voids. The \vf\ analysis clearly shows the prominence of the voids in the LSS, 
not hindered by smoothing of the over-dense regions, and it reveals the image 
of a void-filled universe, where large voids are a common feature.

The consistency in the void image between \iras\ and optically selected 
galaxies suggests that galaxies of different types delineate equally well the 
observed voids. Therefore galaxy biasing is an unlikely mechanism for 
explaining the observed voids in redshift surveys. Comparison with the 
recovered mass distribution further suggests that the observed voids in the 
galaxy distribution correspond well to under-dense regions in the mass 
distribution. This confirms the gravitational origin of the voids \citep{pi93}.

\acknowledgments
We would like to thank Luiz da~Costa for providing us with the SSRS2 data and 
for numerous helpful discussions.
    
%  ****************************************
%  *            Bibliography              *
%  ****************************************

\clearpage

\begin{table*}
{\scriptsize
%{\scriptsize \sc
%2e \begin{tabular}{@{}rccD{.}{.}{-1}D{.}{.}{-1}rrccl@{}}
\begin{tabular}{@{}rccccrrccl@{}}
%\hline \hline
 & Confidence & Equivalent & \centercol{Total} & \multicolumn{3}{c}{Location of Center} & Void & Largest & \\
 & Level & Diameter & \centercol{Volume} & \multicolumn{3}{c}{(Equatorial Coordinates)} & Under- & Sphere's & \\
\cline{5-7}
 & $p$ & {\rm [\hmpc]} & \centercol{\rm [$h^{-3}\:\mbox{KMpc}^3$]} & \centercol{$r$} & \centercol{$\alpha$} & \centercol{$\delta$} & density & Fraction & Identification \\
 & (1) & (2) & \centercol{(3)} & \centercol{(4)} & \centercol{(5)} & \centercol{(6)} & (7) & (8) & \hspace{0.62cm} (9) \\
\hline
% N   --------stat   --d-   ----vol--   ------r--   -right-ascension-   ----declination------    dens   frac     grid  large
  1 & $  >   $0.99 & 54.3 & \phn 84.9 & \phn 85.7 & $ 1\hour 33\Min $ & $-16\arcdeg45\arcmin$ & -0.89 & 0.25 & \\ % 31    34.3 
  2 & $  >   $0.99 & 56.2 & \phn 93.9 & \phn 99.7 & $ 3\phmh 34\phmm$ & $-28\phmdg 50\phmmin$ & -0.87 & 0.20 & dC88 void~4 \\ % 37    33.0 
  3 & $  >   $0.99 & 60.8 &     119.0 &     107.2 & $22\phmh 25\phmm$ & $-14\phmdg 46\phmmin$ & -0.93 & 0.25 & \\ % 39    38.5 
  4 & $  >   $0.99 & 35.6 & \phn 24.0 & \phn 66.7 & $21\phmh 43\phmm$ & $-14\phmdg 40\phmmin$ & -0.91 & 0.37 & \\ % 43    25.5 
  5 & $  >   $0.99 & 34.8 & \phn 22.6 & \phn 53.0 & $23\phmh 48\phmm$ & $-24\phmdg 39\phmmin$ & -0.94 & 0.39 & dC88 void~3 (Sculptor) \\ % 43    25.5 
  6 & $  >   $0.99 & 32.0 & \phn 17.4 & \phn 56.5 & $ 3\phmh 56\phmm$ & $-20\phmdg 11\phmmin$ & -0.92 & 0.47 & \\ % 43    25.0 
  7 & $  >   $0.99 & 25.5 &\phn\phn 8.8&\phn 77.2 & $ 3\phmh 17\phmm$ & $-11\phmdg 40\phmmin$ & -0.91 & 0.71 & \\ % 47    22.6 
  8 & $  >   $0.99 & 27.8 & \phn 11.4 & \phn 83.9 & $23\phmh 20\phmm$ & $-12\phmdg 32\phmmin$ & -0.95 & 0.59 & \\ % 49    23.4 
  9 & $\phmgt$0.99 & 39.0 & \phn 31.1 &     114.6 & $ 3\phmh 06\phmm$ & $-13\phmdg 47\phmmin$ & -0.86 & 0.54 & \\ % 51    31.7 
 10 & $\phmgt$0.95 & 34.8 & \phn 22.4 &     104.7 & $ 0\phmh 26\phmm$ & $ -9\phmdg 17\phmmin$ & -0.69 & 0.39 & \\ % 55    25.5 
\vspace{6pt}
 11 & $\phmgt$0.95 & 42.9 & \phn 41.5 &     112.8 & $ 0\phmh 21\phmm$ & $-29\phmdg 43\phmmin$ & -0.88 & 0.31 & \\ % 55    28.9 
 12 & $\phmgt$0.72 & 25.0 &\phn\phn 8.1&\phn 74.8 & $23\hour 03\Min $ & $-32\arcdeg35\arcmin$ & -0.97 & 0.37 & \\ % 60    17.9 
 13 & $\phmgt$0.72 & 22.1 &\phn\phn 5.8&\phn 31.0 & $ 1\phmh 23\phmm$ & $-19\phmdg 36\phmmin$ & -1.00 & 0.55 & dC88 void~1 \\ % 60    18.2 
 14 & $\phmgt$0.53 & 21.3 &\phn\phn 5.2&\phn 87.2 & $21\phmh 28\phmm$ & $-29\phmdg 28\phmmin$ & -1.00 & 0.62 & \\ % 63    18.2 
 15 & $\phmgt$0.53 & 27.3 &\phn  10.7 &     116.1 & $21\phmh 24\phmm$ & $-33\phmdg 17\phmmin$ & -0.74 & 0.86 & \\ % 63    26.0 
 16 & $\phmgt$0.36 & 20.3 &\phn\phn 4.4&\phn 36.5 & $21\phmh 43\phmm$ & $-18\phmdg 41\phmmin$ & -1.00 & 0.52 & \\ % 66    16.1 
 17 & $\phmgt$0.27 & 19.0 &\phn\phn 3.7&\phn 32.1 & $ 3\phmh 42\phmm$ & $-21\phmdg 21\phmmin$ & -1.00 & 0.62 & \\ % 69    16.1 
 18 & $\phmgt$0.27 & 21.1 &\phn\phn 4.9&\phn 85.9 & $ 4\phmh 18\phmm$ & $ -8\phmdg 42\phmmin$ & -1.00 & 0.50 & \\ % 69    16.6 
%\hline
\end{tabular}
}
\caption[Voids in the SSRS2 survey: locations and properties]
{Locations \& properties of the voids in the SSRS2 survey.}
\label{ssrs2-table}
\end{table*}
\clearpage

\clearpage

\clearpage

\clearpage

\clearpage

\notetoeditor{The following two figures should appear on facing pages, to
allow the reader easy comparison between them.}
\clearpage

\clearpage

\clearpage

\clearpage

\clearpage

\notetoeditor{The following two figures should appear on facing pages, to
allow the reader to examine all six slices one after the other. Note also that
this is ONE figure, extending on TWO pages.}
\begin{figure*}
\centering
%graphicx: \includegraphics[width=0.82\linewidth]{s2_1.ps}
\epsfig{file=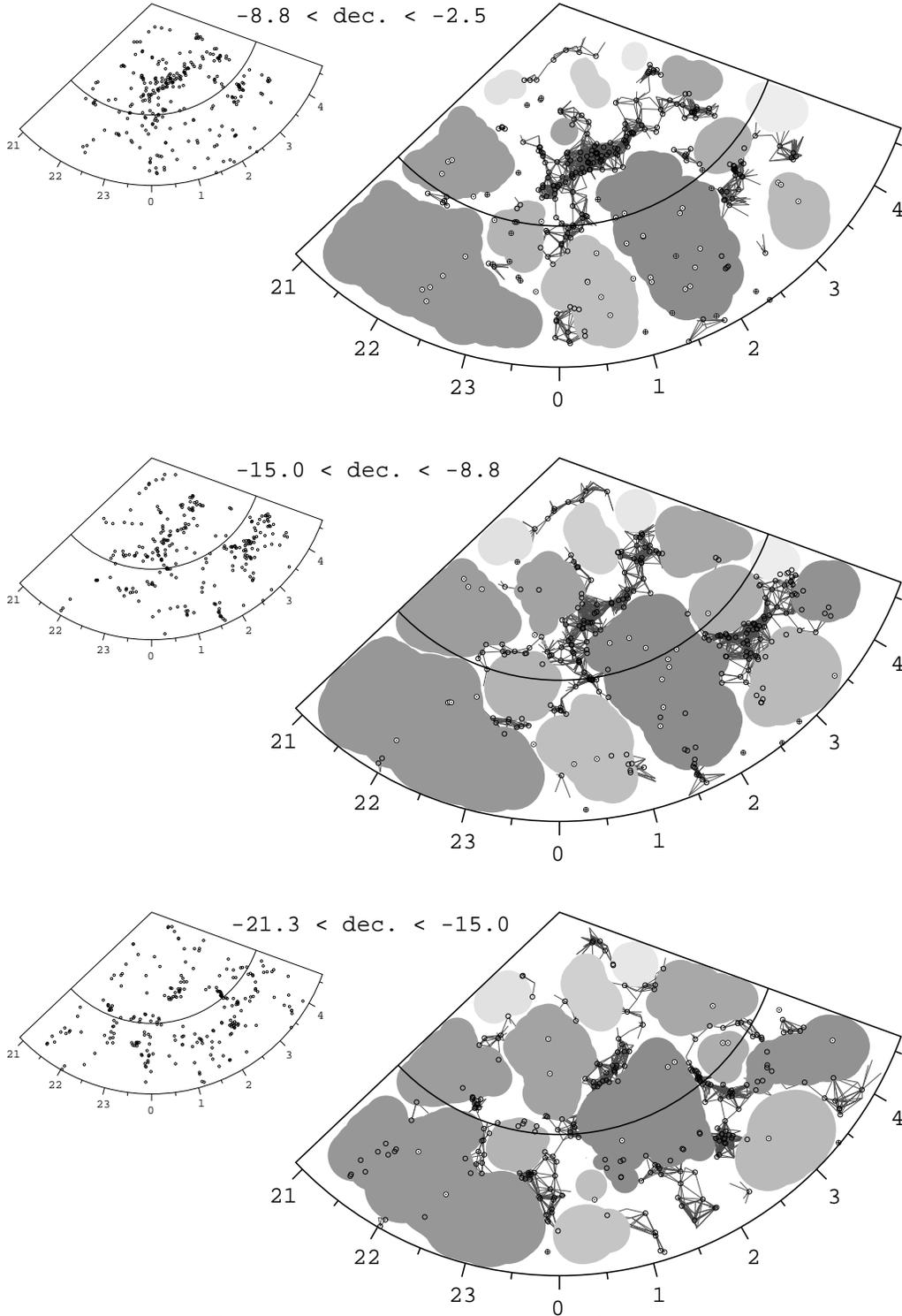, width=0.82\linewidth}
\caption[The SSRS2 - part 1]
{LSS in the SSRS2. The survey spans the declination range 
$ -40\arcdeg < \delta < -2\fdg5 $, which we have split to six $6\fdg25$-wide 
slices. Our sample extends to $ \rmax = 130 \hmpc $, and the intermediate arc 
marks $ \rvol = 79.5 \hmpc $ (\emph{Continued on the following page}).}
\label{ssrs2-fig}
\end{figure*}
\clearpage

\addtocounter{figure}{-1}
\begin{figure*}
\centering
%graphicx: \includegraphics[width=0.82\linewidth]{s2_2.ps}
\epsfig{file=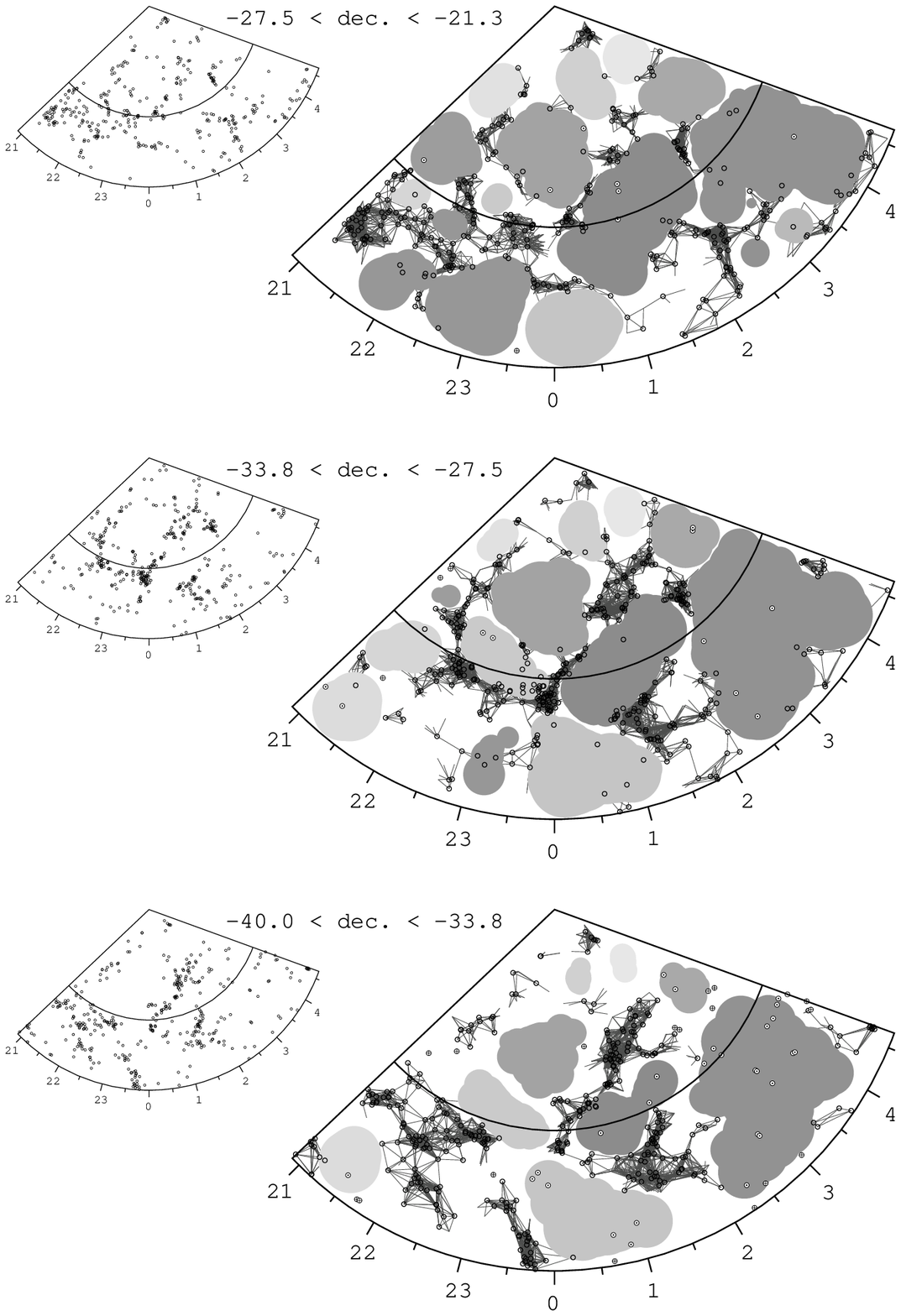, width=0.82\linewidth}
\caption[The SSRS2 - part 2]
{(\emph{Continued from previous page})
 Voids (grey areas), walls (dark lines) and galaxies 
 (wall galaxies: `$\circ$'; 
 void galaxies: `$\odot$'; 
 field galaxies: `$\oplus$').}
%\label{ssrs2-fig}
\end{figure*}
\clearpage

\begin{figure}
\centering
%graphicx: \includegraphics[width=\linewidth]{ssrs2.ra.ps}
\plotone{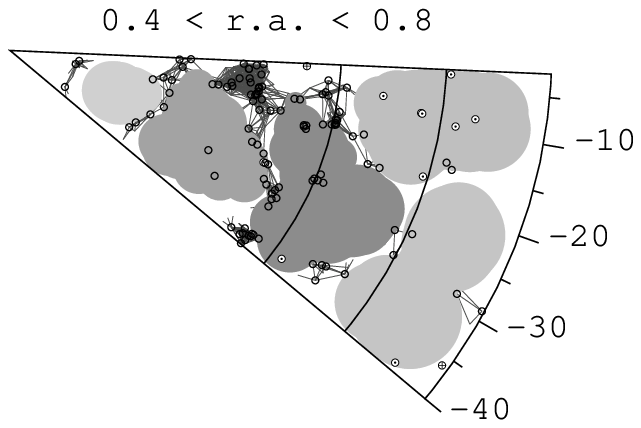}
\caption[SSRS2 declination range]
{A constant right-ascension slice of the SSRS2, demonstrating the 
narrow ($37\fdg5$) declination range of this survey. Most of the voids in 
this image are limited by the survey's boundary.}
\label{ssrs2-ra-fig}
\end{figure}
\clearpage

\begin{figure*}
\centering
%graphicx: \includegraphics[width=0.42\linewidth]{ssrs2.sgp_zs.ps}
%graphicx: \hspace{7mm}
%graphicx: \includegraphics[width=0.42\linewidth]{iras.sgp_zs.ps}
\plottwo{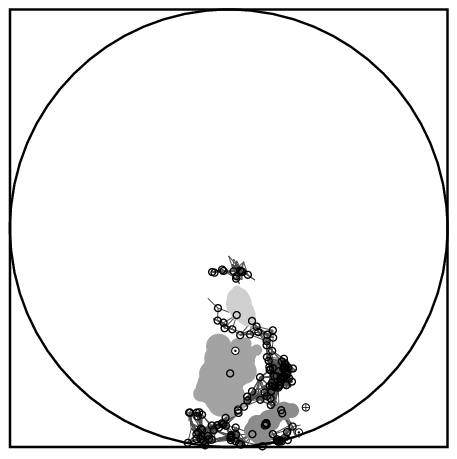}{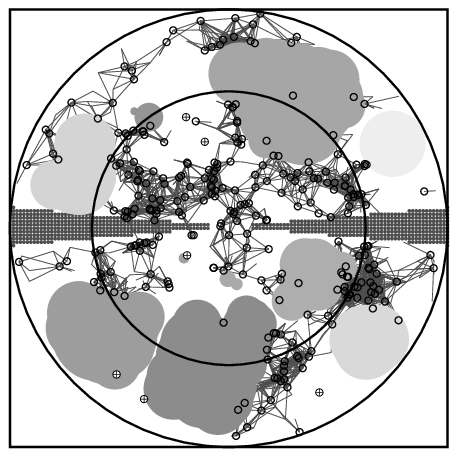}
\caption[\iras\ and SSRS2 voids in the SG plane]
{Voids in the SG plane: here we compare the redshift-space voids of 
the SSRS2 (\emph{left}) and the \iras\ (\emph{right}). The denser sampling of 
the SSRS2 is evident. Similar voids are found in the overlapping regions of 
the two surveys.}
\label{iras-ssrs2-fig}
\end{figure*}

\end{document}